\documentstyle[prb,aps,rotate,array,multicol]{revtex}
\input{epsf}
\newcommand{\lrule}{ \end{multicols} \noindent
  \rule{0.5\textwidth}{0.1mm}\rule{0.1mm}{3pt}\newline }
\newcommand{\rrule}{ \noindent \parbox{\textwidth}{
  \hfill\rule[-3pt]{0.1mm}{3pt}\rule{0.5\textwidth}{0.1mm}}
  \begin{multicols}{2}\noindent }

\begin{document}

\draft


\title{Josephson current between chiral superconductors}

\author{Yu.~S.~Barash$^{1,2}$ and A.~M.~Bobkov$^{2}$}
\address{$^1$Center for Electronic Correlations and Magnetism, Institute of Physics,
University of Augsburg, D-86135 Augsburg, Germany\\
$^2$ P.N.~Lebedev Physical Institute, Leninsky Prospect 53,
 Moscow 119991, Russia}
\author{M.~Fogelstr\"om}
\address{Institute of Theoretical Physics, Chalmers University of Technology
and G\"oteborg University,
S-41296 G\"oteborg, Sweden}




\maketitle


\begin{abstract}
We study chiral interface Andreev bound states and their influence
on the Josephson current between clean superconductors. Possible
examples are superconducting \mbox{Sr$_2$RuO$_4$} and the
$B$-phase of the heavy-fermion superconductor \mbox{UPt$_3$}. We
show that, under certain conditions, the low-energy chiral surface
states enhance the critical current of symmetric tunnel junctions
at low temperatures. The enhancement is substantially more
pronounced in quantum point contacts. In classical junctions
dispersive chiral states result in a logarithmic dependence of the
critical current. This logarithmic behavior contains the
temperature, the barrier transparency and  the broadening of the
bound states, and depends on the detailed relation between these
parameters. The Josephson current through the domain wall doesn't
acquire this logarithmic enhancement, although the contribution
from the bound states is important in this case as well.
\end{abstract}




\begin{multicols}{2}
\section{Introduction}

Superconductors in a state with both time-reversal symmetry and
parity with respect to the inversion of a crystal axis broken are
sometimes named chiral superconductors. The "chirality" can be
associated with, for example, a nonzero projection of the orbital
angular momentum, $\bbox{l}$, of Cooper pairs along a axis
$\bbox{z}$, i. e. $l_z\ne 0$. Superconducting \mbox{Sr$_2$RuO$_4$}
as well as the heavy-fermion superconductor \mbox{UPt$_3$}\, in
its $B$-phase (i.e. the low-temperature and low-field phase) are
leading candidates for chiral superconductors of this type.
Another possible reason for a nonzero chirality of a
superconducting state to appear is a complex admixture of two
pairing channels belonging to different irreducible
representations. For two-dimensional chiral superconductors there
is a topological invariant associated with the
chirality\cite{vol97}.

One of the important features of superconductors is the presence
of surface or interface Andreev bound states. They can take place
in the vicinity of surfaces or interfaces if the order parameter
varies in space or takes different values for incoming and
outgoing momentum directions along a quasiclassical trajectory
travelled by a quasiparticle in a reflection or a transmission
event. Surface and interface quasiparticle states arising in
superconductors already in the absence of the magnetic field, have
been studied theoretically for many years (for example,
see\cite{barsagi72,buchholtz81}). They have attracted much
attention in investigating high temperature superconductivity. For
instance, the zero-energy surface states in $d$-wave
superconductors appear near an impenetrable smooth surface as a
consequence of a sign change of the order parameter along
quasiparticle trajectories connected by the scattering off the
surface. Zero-energy bound states lead at low temperatures to a
zero-bias conductance peak of N-I-D (normal metal - isolating
barrier - $d$-wave superconductor) tunnel
junctions\cite{hu94,tan95,fog97,bbs97}, which was observed
in experiments\cite{cov97,alf97,ek97,ap98,alf981,alf982,sin98,wei981,wei982,%
apr99,deutch99,cucolo00}. Low-temperature anomalies, originating from
the zero-energy bound states, are also seen in the Josephson
critical current\cite{tk96,bbr96,rb98,ovs00,il01} and in the
penetration depth\cite{wal98,alf981,bkk00,car01}. The strength of these effects depend
on the broadening of the bound states by surface
roughness\cite{bbr96,wal98,bkk00} and bulk
impurities\cite{pbbi99}, as well as on their shift from zero on
account of junction transparency\cite{tk96,rb98,bar00}.

Zero-energy surface states are dispersionless states, while
Andreev bound states with nonzero energy are dispersive: the
energy depends on a quasiparticles momentum direction. Bound states
at energies on the order of $\Delta(\bbox{p}_f)$ can manifest
themselves, for example, in the current-voltage characteristics of a
junction along with the low-energy bound states\cite{bbs97,bs97}.
They do not, however, modify noticeably equilibrium
characteristics like the Josephson critical current or the
penetration depth. In contrast, the zero- and/or low-energy states
can have strong influence on the equilibrium quantities at low
temperatures.

Chiral surface states\cite{vol97,honer98,matsum99} make up a
special type of Andreev bound states. They form dispersive
branches, which can cross the zero level only for isolated
quasiparticle momentum directions. The spontaneous surface current
in chiral superconductors\cite{vol97,matsum99}, the spectrum of
the surface chiral bound states at an impenetrable
wall\cite{honer98,matsum99}, and their contributions to
the density of states, the conductance peak of the tunnel
junctions\cite{honer98,matsum99,tan99,fog00}, and the
low-temperature penetration depth\cite{morinari00}, have already
been studied theoretically in literature. There are, however,
various open problems in the field, part of which is addressed in
the present paper. We report on theoretical results for the
spectra of chiral surface and interface states and their
contribution to the Josephson critical current of junctions
between chiral superconductors. In experiments to date the
Josephson effect has been studied only in junctions between
\mbox{Sr$_2$RuO$_4$} (or \mbox{UPt$_3$}) and a conventional
superconductor\cite{jin00,sumiyama98}. Our results show that
extending experimental investigations of the Josephson critical
current in symmetric junctions of the ruthenates (or
\mbox{UPt$_3$}) would sensitively probe the chiral nature of these
superconductors.

Our numerical approach is to solve the quasiclassical transport
equation and to determine the profile of the order parameter
self-consistently. Our analytical results describe spectra of the
chiral states at interfaces of any transparency assuming spatially
constant (non-self consistent) order parameters. These
calculations are improved further for low-energy states in the
tunnel-junction limit, where the surface pair-breaking may be
taken into account analytically as well. As is demonstrated below,
the low-energy parts of the chiral branches of the surface Andreev
bound state spectra can dominate and strengthen the Josephson
critical current in symmetric tunnel junctions at sufficiently low
temperatures. The low-temperature enhancement of the Josephson
critical current, $j_c$, is substantially more pronounced for
quantum point contacts. It is of similar origin as the
low-temperature anomaly in $j_c$ of junctions between $d$-wave
superconductors\cite{tk96,bbr96}. In classical junctions the
dispersion of the chiral states strongly modifies the anomalous
low-temperature behavior of $j_c$, leading eventually to an
additional logarithmic factor, which can be large under certain
conditions: if the broadening of the bound states and the
transparency of the junction are sufficiently small. The
zero-temperature value of $j_c$ is then determined by a presumably
small broadening of the bound states or/and a small junction
transparency that cuts off the logarithmic divergence as
$T\rightarrow0$.

We also study the Josephson current through a domain wall,
considering a junction between identical equally oriented
superconductors with opposite chiralities. We demonstrate the
crucial role of the Andreev interface states for obtaining a finite
Josephson current in the system with the crystal $c$-axis parallel
to the interface. Disregarding the presence of surface states and
assuming spatially constant order parameters, one can apply at the
interface the bulk expression for the Green's function. Then the
tunnel Josephson critical current across the domain wall vanishes.
If the $c$-axis is along the interface normal and the interface
itself is symmetric with respect to rotations around the normal,
then the dc Josephson current actually vanishes, which is known
for a long time for junctions between superconductors with
opposite projections of the angular momentum of Cooper
pairs\cite{amb74,mrs88}. For the $c$-axis parallel to the surface,
vanishing current is, however, a shortcoming of the
oversimplified approach. We show, that the interface influence
results in this case both in Andreev bound states and in a
finite Josephson current close to the conventional value.

We perform our analytical and numerical calculations, first, for a
quasi two-dimensional tetragonal superconductor, choosing several
particular basis functions, which are considered as candidates for
the superconducting pairing in \mbox{Sr$_2$RuO$_4$}. Possible candidates
for the order parameter in \mbox{Sr$_2$RuO$_4$} are still intensively discussed
\cite{rice95,mazin97,luke98,maeno98,miyake99,sigzhit99,maki99,sidis99,%
has00,fog00,graf00,maki00,won00,kealey00,bonalde00,matsui01,eschrig01,lupien01,ishida01,%
tanatar012,tanatar011,tewordt01,joynt01,annett01}. For our study
we take three types of two-component triplet order parameters
forming the $p$-wave, the $f_{xy}$-wave and the $f_{x^2-y^2}$
superconducting $(1,\pm i)$ phases.

We also present analytical results, demonstrating that analogous
chiral interface bound states and low-temperature enhancement
in the Josephson current take place in a three-dimensional
hexagonal superconductor like \mbox{UPt$_3$}. For many years the
two-component triplet $E_{2u}$-representation and singlet
$E_{1g}$-representation order parameters were considered as
the probable candidates for the pairing state in
\mbox{UPt$_3$}\cite{sigueda,sauls,heffner96,goll99}. In particular,
$(1,\pm i)$ superconducting phases are supposed to form at low
temperatures and in weak magnetic fields. The analysis of
experimental data now available has led to
a presumably definite choice in favor of the $E_{2u}$ type of
pairing in \mbox{UPt$_3$}\cite{GYS00,hux00,min01}. The so-called "3D"
model for the order parameter in tetragonal \mbox{Sr$_2$RuO$_4$} is based
on an analogous type of pairing\cite{maki99}. We examine
pairing both in the $E_{2u}$ and in the $E_{1g}$ representation for three-dimensional
superconductors and find conditions, when only one of them
results in the anomaly of the Josephson current we study below.

\section{Chiral Surface and Interface Quasiparticle States}

\subsection{Chiral order parameters}

Let a superconductor breaking time-reversal symmetry
occupy a half-space with a surface (or interface) normal along the
$x$-axis. The superconductor may be both triplet or singlet.
Triplet order parameters, $\hat{\Delta}(\bbox{p}_f)=
\bbox{d}(\bbox{p}_f)\!\cdot\!\hat{\bbox\sigma}\, i\hat{\sigma}_y$,
considered below have only the component
$d_z(\bbox{p}_f)\equiv\Delta(\bbox{p}_f)$ nonzero.

Unconventional order parameters are in general quite sensitive to
any inhomogeneity in the superconductor that gives rise to
quasiparticle scattering and this in particular to a boundary. We
represent a chiral order parameter near a surface or interface as
$\Delta(\bbox{p}_f,x)=\Bigl(
\Delta_1(\bbox{p}_f,x)e^{i\phi_1(x)}+i\chi\Delta_2(\bbox{p}_f,x)
e^{i\phi_2(x)}\Bigr)e^{i\Phi}=|\Delta(\bbox{p}_f,x)|
e^{i\phi(\bbox{p}_f,x)}$, where the two real components
$\Delta_1(\bbox{p}_f,x)$ and $\Delta_2(\bbox{p}_f,x)$ have
presumably opposite parities with respect to the inversion of the
$x$-axis. The parameter $\chi=\pm 1$ corresponds to two possible
values of "chirality". The phases $\phi_1(x)$ and $\phi_2(x)$
vanish in the bulk but may be nonzero in vicinity of the boundary.

Assuming a model with a single cylindrical Fermi surface, the two components of the
$p$-wave order parameter in a tetragonal superconductor with the crystal
$x_0$-axis along the surface normal can be taken in the bulk as
$\Delta^p_{1,b}(\bbox{p}_f)=\Delta^p_{b}\cos\varphi$ and $\Delta^p_{2,b}(
\bbox{p}_f)=\Delta^p_{b}\sin\varphi$. Here and below $\varphi$ is the
angle which the vector $(p_{f,x},p_{f,y},0)$ makes with the $x$-axis.
For the $f_{xy}$-wave order parameter in the bulk, one has
$\Delta^{f_{xy}}_{1,b}(\bbox{p}_f)=\Delta^{f_{xy}}_{b}\sin2\varphi\cos\varphi$
and $\Delta^{f_{xy}}_{2,b}(\bbox{p}_f)=\Delta^{f_{xy}}_{b}\sin2\varphi
\sin\varphi$.\ For the bulk $f_{x^2-y^2}$-wave order parameter:
$\Delta^{f_{x^2-y^2}}_{1,b}(\bbox{p}_f)=\Delta^{f_{x^2-y^2}}_{b}\cos2\varphi
\cos\varphi$ and $\Delta^{f_{x^2-y^2}}_{2,b}(\bbox{p}_f)=\Delta^{f_{x^2-y^2}}_{b}
\cos2\varphi\sin\varphi$.

If $\varphi$ describes the incoming momentum direction specular
quasiparticle reflection from the surface gives $(\pi-\varphi)$
for the outgoing momentum. This corresponds to the inversion of
the $x$-axis. One can easily see that order parameter components
$\Delta_{1}(\bbox{p}_f)$ and $\Delta_{2}(\bbox{p}_f)$ have
opposite parities with respect to the inversion of the $x$-axis in
all cases considered. Since the total order parameter represents a
complex mixture of $\Delta_{1}(\bbox{p}_f)$ and
$\Delta_{2}(\bbox{p}_f)$, we get, indeed, chiral order parameters
describing superconducting states with both time-reversal symmetry
and parity with respect to the inversion of the $x$-axis broken.
An additional feature, which makes our analytical results
comparatively simple, is that in all cases we consider
$|\Delta_b(\bbox{p}_f)|$ doesn't change in the reflection event
(if $\phi_{1,2}(x)=0$, then $|\Delta(\bbox{p}_f,x)|$ doesn't
change).

\begin{figure}
\centerline{
\rotate[r]{\epsfxsize=0.49\textwidth{\epsfbox{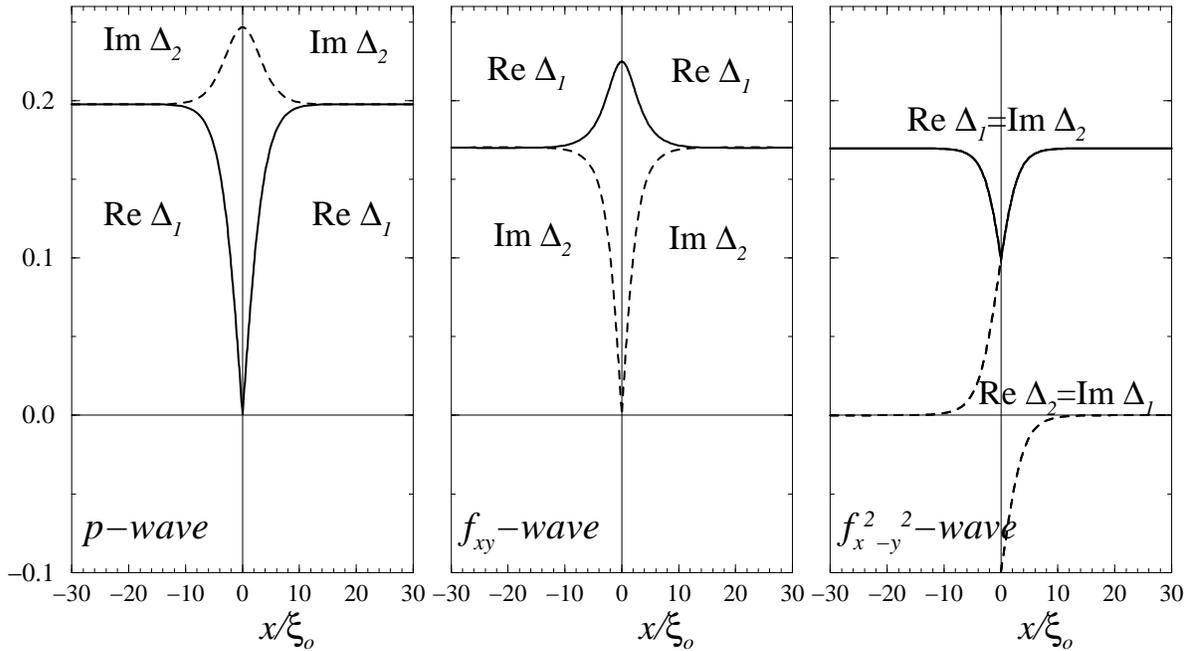}}}}
\caption[]{ The spatial profiles of the order-parameter components
$\Delta_{1,2}(x)$ (in units $2\pi T_c$). } \label{OrderParameters}
\end{figure}

In Figure \ref{OrderParameters} we show order-parameter
profiles at an impenetrable barrier separating a
$\Delta_{1}(\bbox{p}_f,x)e^{i\phi_1(x)}+i\Delta_{2}(\bbox{p}_f,x)
e^{i\phi_2(x)}$ (left side) and a $\Delta_{1}(\bbox{p}_f,x)e^{i\phi_1(x)}-
i\Delta_{2}(\bbox{p}_f,x)e^{i\phi_2(x)}$
(right side) superconductor. The temperature is 0.001 $T_c$.
The order parameter components are represented in the forms
$\Delta_{1}(\bbox{p}_f,x)e^{i\phi_1(x)}\equiv\Delta_{1}(x)\psi_{2}(\bbox{p}_f)
$, $\pm i\Delta_{2}(\bbox{p}_f,x)e^{i\phi_2(x)}\equiv\Delta^{l(r)}_{2}(x)
\psi_{2}(\bbox{p}_f)$, which define the profiles $\Delta^{l(r)}_{1,2}(x)$
displayed in Fig.\ \ref{OrderParameters}. They are computed
using a single cylindrical Fermi surface and the pairing potential
$V(\bbox{p},\bbox{p}')=V_0[\psi_1(\bbox{p})
\psi_1(\bbox{p}')+\psi_2(\bbox{p})\psi_2(\bbox{p}')]$ in a
factorized form with the simple, normalized, basis functions
\begin{equation}
(\psi_1(\bbox{p}_f),\psi_2(\bbox{p}_f))=
\bigg\lbrace\begin{array}{ll}
\sqrt{2}\, (\cos\varphi,\sin\varphi)&\mbox{(p-wave)}\\
2\, (\cos\varphi,\sin\varphi)\sin 2\varphi&\,
\mbox{($f_{xy}$-wave)} \enspace .
\end{array}
\label{waves}
\end{equation}

\end{multicols}

\begin{multicols}{2}
For the $p$-wave case our self-consistent numerical results are in
agreement with those represented in\cite{matsum99}. One should
note the vanishing $\Delta_1(x,\bbox{p}_f)$ at the surface and, as a
consequence, the increase in $\Delta_2(0,\bbox{p}_f)$ as compared
to its bulk value. An inhomogeneous profile
$\Delta_1(x,\bbox{p}_f)$ near the surface slightly modifies the
surface value $\Delta_2(0,\bbox{p}_f)$ as well\cite{matsum99}. If
the crystal axis $x_0$ makes a finite angle $\alpha$ in the
$xy$-plane with the surface normal, for the p-wave superconductor
there is no dependence on $\alpha$. The surface state of an f-wave
superconductor, however, will depend on $\alpha$. The surface
profiles of the $f$-wave order parameters with angles $\alpha=0$
and $\alpha=45^\circ$ are shown in Fig. \ref{OrderParameters} and
correspond to the most extreme cases, $f_{xy}$ and $f_{x^2-y^2}$
respectively. We note, that the $f_{xy}$-wave order parameter
behaves quite similar to the $p$-wave case near a surface.
For $p$-wave and $f_{xy}$-wave pairings the phases $\phi_1(x)$ and $\phi_2(x)$ obey
$\phi_1(x)=\phi_2(x)\equiv0$, while for the $f_{x^2-y^2}$-wave order
parameter $\phi_1(x)=-\phi_2(x)\ne 0$. For all pairing states we
see that the order parameter has healed after roughly 15 $\xi_0, (
\xi_0=v_{\!f}/2\pi T_c)$. If instead the right superconductor
would have been calculated with the same chirality as the left
one, the order parameter would change as
$(\Delta_{1}(\bbox{p}_f,x),\Delta_{2}(\bbox{p}_f,x))
\rightarrow(\Delta^*_{1}(\bbox{p}_f,x),-\Delta^*_{2}(\bbox{p}_f,x))$
on the right side. When we continue to study the Josephson
coupling of two chiral superconductors we shall assume that the
Josephson currents generated are much smaller than the
bulk critical current and do not affect the
spatial dependence of the order parameter or the surface currents
(see below).

For three-dimensional models of chiral superconducting states we
assume, for simplicity, a spherical Fermi surface and choose bulk
order parameters in the form:
$\Delta^t_{z,b}(\bbox{p}_f)=\Delta^t_{b}\cos\theta\sin^2\theta
e^{i( 2\chi\varphi+\Phi)}$ for the $E_{2u}$-representation
(triplet pairing) and
$\Delta^s_{b}(\bbox{p}_f)=\Delta^s_{b}\cos\theta\sin\theta
e^{i(\chi\varphi+\Phi)}$ for the $E_{1g}$ (singlet pairing).
Hence, for the $E_{2u}$ and the $E_{1g}$ pairings respectively we
get
\lrule
\begin{equation}
\begin{array}{ccc}
{\left\{\begin{array}{c} \Delta^t_{1,b}(\bbox{p}_f)=\Delta^t_{b}
\cos\theta\sin^2\theta\cos 2\varphi \enspace ,\\
\Delta^t_{2,b}(\bbox{p}_f)=\Delta^t_{b}
\cos\theta\sin^2\theta\sin 2\varphi \enspace ,\\
\end{array}\right.}
&\protect\hspace*{1.8cm} & {\left\{\begin{array}{c}
\Delta^s_{1,b}(\bbox{p}_f)=\Delta^s_{b}
\cos\theta\sin\theta\cos\varphi \enspace ,\\
\Delta^s_{2,b}(\bbox{p}_f)=\Delta^s_{b}
\cos\theta\sin\theta\sin\varphi \enspace .\\
\end{array}\right.}\\
\end{array}
\label{bts}
\end{equation}
Here $\theta$ and $\varphi$ are standard angles in a spherical
system of coordinates in momentum space with the $x$-axis along
the surface normal. $\Phi$ is a spatially constant part of the
complex phase of the order parameter. Order parameter components
$\Delta_2^t(\bbox{p}_f,x)$ and $\Delta_1^s(\bbox{p}_f,x)$ vanish
on an impenetrable surface.

\subsection{Interface chiral bound states}

Consider a junction between identical equally oriented chiral
superconductors. Solving Eqs.(\ref{pe})-(\ref{asr}) of the Appendix
with spatially constant anisotropic order
parameters on both sides of the interface we find comparatively
simple analytical expressions for the energies of chiral interface
Andreev bound states. If $\tilde\chi\equiv\chi^l\chi^r=1$, i.e.
the superconductors have identical chiralities, we get the
following spectrum for two-dimensional $p$-wave and
$f_{x^2-y^2}$-wave order parameters as well as for the
three-dimensional singlet $E_{1g}$-pairing:
\begin{equation}
\varepsilon^{p,f_{x^2-y^2},s}_B(\bbox{p}_f)=\pm
\left|\Delta_b^{p,f_{x^2-y^2},s}(\bbox{p}_f)\right|\left[ {\cal
R}\sin^2\varphi+{\cal D}\cos^2 \left ( \frac{\Phi}{2}\right )
\right]^{1/2} \enspace . \label{ecds}
\end{equation}
For the $E_{2u}$- and the $f_{xy}$-pairing we obtain:
\begin{equation}
\varepsilon^t_B(\bbox{p}_f)=\pm
\left|\Delta_b^{t}(\bbox{p}_f)\right|\left[ {\cal R}\cos^2
2\varphi+{\cal D}\cos^2 \left ( \frac{\Phi}{2}\right )
\right]^{1/2} \enspace , \label{ecdt}
\end{equation}
\begin{equation}
\varepsilon^{f_{xy}}_B(\bbox{p}_f)=\pm
\left|\Delta_b^{f_{xy}}(\bbox{p}_f)\right|\left[ {\cal
R}\cos^2\varphi+{\cal D}\cos^2 \left ( \frac{\Phi}{2}\right )
\right]^{1/2} \enspace . \label{ecdf1}
\end{equation}
Here ${\cal D}$ and ${\cal R}$ are transmission and reflection
coefficients of the interface barrier which in general depend
on the quasiparticle momentum direction. The phase difference
$\Phi=\Phi^r-\Phi^l$,
where $\Phi^{r,l}$ are the phases of the order parameters in the
right and the left superconductors, is separated from the intrinsic
chiral phases, $\varphi$, as defined in the preceding subsection. The
spectra in Eqs.(\ref{ecds})-(\ref{ecdf1}) depend on momentum
directions both explicitly and via ${\cal D}$ and ${\cal R}$. They
significantly differ from the bound state energies in junctions
between conventional
superconductors\cite{fur90,fur191,been191,been291}
$\varepsilon_B(\bbox{p}_f)=\pm|\Delta|\sqrt{1-{\cal
D}\sin^2\frac{\displaystyle \Phi}{\displaystyle 2}}$.

Positive and negative branches in Eqs.(\ref{ecds})-(\ref{ecdf1})
always touch on the Fermi surface (on the zero-energy level) for
momentum directions along possible nodes of
$|\Delta(\bbox{p}_f)|$. In addition, if the phase difference
$\Phi$ is equal to $\pi$ (or for any $\Phi$, but in the limit of
an impenetrable wall ${\cal D}=0$), the branches described by
Eqs.(\ref{ecds}) and (\ref{ecdt}), acquire new crossing points (or
lines) lying on the Fermi surface. These additional momentum
directions are at $\varphi=0,\pi$ for the $p$-wave, the
$f_{x^2-y^2}$-wave and the three-dimensional $E_{1g}$-pairing,
while at $\varphi=\pm \pi/4, \pm 3\pi/4$ for the $E_{2u}$-pairing.
In accordance with Eq.(\ref{ecdf1}) for the $f_{xy}$-pairing the
spectrum of the bound states takes zero value only for directions
along the nodes of the order parameter, i.e at $\varphi=0,\pi,\pm
\pi/2$. At the glancing trajectories $\varphi=\pm\pi/2$ the nodes
in the spectrum have a higher multiplicity.

It is worth to compare interface states in
Eqs.(\ref{ecds})-(\ref{ecdf1}) with the respective surface chiral
branches at an impenetrable wall:
\begin{equation}
\varepsilon^{p,s,f_{x^2-y^2}}_B(\bbox{p}_f)=\chi\,{\rm sgn}(x)
\left|\Delta^{p,s,f_{x^2-y^2}}_{b}(\bbox{p}_f)\right|\sin\varphi \enspace ,
\label{ebpc}
\end{equation}
\begin{equation}
\begin{array}{lr}
\varepsilon^{f_{xy}}_B(\bbox{p}_f)=-\chi\,{\rm sgn}(x)
\Delta^{f_{xy}}_b\sin2\varphi\cos\varphi \enspace ,\qquad &
\varepsilon^t_B(\bbox{p}_f)=-\chi\,{\rm sgn}(xp_y)
|\Delta^t_{b}(\bbox{p}_f)|\cos2\varphi
\enspace .\\
\end{array}
\label{ebsc}
\end{equation}
\rrule
One can see, that finite transmission and $\Phi\ne\pi$ result in
an effective "repulsion" and a reconnection of the branches from
two sides of the barrier plane.
Chiral states on an impenetrable surface of the $p$-wave
superconductor were obtained earlier in Refs.\ \onlinecite{honer98,matsum99}.
In the particular case $\Phi=\pi$ Eqs.(\ref{ebpc}) and (\ref{ebsc}) describe
as well the states on interfaces with arbitrary transparency if compensated
by a factor $\sqrt{\cal{R}}$.

The chiral spectra in Eqs.(\ref{ebpc}) and (\ref{ebsc}) are
antisymmetric with respect to the inversion $p_y\to-p_y$ (or
equivalently $\varphi\to-\varphi$), as well as to the inversion of
the total momentum $\bbox{p}_f\to-\bbox{p}_f$ (or
$\varphi\to\varphi+\pi$). Since at zero temperature the states
with negative (positive) energies are occupied (empty) the chiral
states carry a spontaneous current along the
surface\cite{vol97,matsum99}. This can be considered as a
characteristic feature of chiral states. The current flows near
the surface parallel to the $y$-axis within a thickness the order of
the coherence length $\xi_0$. With the model of a constant order
parameters we find at zero temperature: $j_{y}=-\zeta\chi{\rm
sgn}x\ e{\rm v}_fN_f\Delta_b$, where $\zeta^p=1$,
$\zeta^{f_{x^2-y^2}}=0.5$, $\zeta^{f_{xy}}=-1$,
$\zeta^t\approx0.08$, $\zeta^s=0.25$. The induced screening
supercurrents decay on the order of the penetration depth,
$\lambda$, so that the total surface current is actually zero.

As it follows from Eqs.(\ref{ebpc}) and (\ref{ebsc}), the
quasiparticle surface states on the two sides of the impenetrable wall
are related as
$\varepsilon^l_{B}(\varphi)=-\tilde{\chi}\cdot\varepsilon^r_{B}(\varphi)$,
where $\tilde{\chi}\equiv\chi^l\chi^r$. Hence, if $\tilde\chi=1$,
the spontaneous currents on either side of the interface flow in
opposite directions but with equal magnitude. In the case of finite
transparency the surface chiral states from both sides couple into
interface states. The absence of the total interface current
on the scale $\xi_0$ is accompanied with the recovery of the
symmetry $p_y\to-p_y$ in the spectra in
Eqs.(\ref{ecds})-(\ref{ecdf1}) (unless $\Phi=\pi$ when there is
no coupling between the surface states from two sides). It is
worth noting that at small transparency the local spectral weight
of the states at $x=-0$ (or $x=+0$) is still asymmetric (see Fig.
\ref{Densityofstates} below).

If identical, equally oriented, massive superconductors on each
side of the interface have opposite chiralities the junction can
be considered as an isolating domain wall between two degenerate
chiral phases of the bulk superconducting state. In this case
$\tilde\chi=-1$ and spontaneous surface currents in the two banks
of the interface flow along the same direction. Then the recovery
of the symmetry due to finite transparency doesn't take place.
Indeed, solving Eqs.(\ref{pe})-(\ref{asr}) of the Appendix with a
model spatially constant anisotropic order parameter, having
opposite chiralities on the two sides of the junction, results in
\lrule
\begin{equation}
\varepsilon^{p,f_{x^2-y^2},s}_B(\bbox{p}_f)=\chi^r|\Delta_b^{p,f_{x^2-y^2},s}(
\bbox{p}_f)|\left(\sqrt{1-{\cal D}\cos^2\frac{\Phi}{2}}\sin\varphi\pm\sqrt{
{\cal D}}\cos\frac{\Phi}{2}\cos\varphi\right)\enspace , \label{ecdsq}
\end{equation}
\begin{equation}
\varepsilon^t_B(\bbox{p}_f)=-\chi^r{\rm sgn}~(p_y)
|\Delta_b^{t}(\bbox{p}_f)|\left(\sqrt{1-{\cal
D}\sin^2\frac{\Phi}{2}}\cos2\varphi\pm\sqrt{{\cal
D}}\sin\frac{\Phi}{2}\sin2\varphi\right)\enspace , \label{ecdtq}
\end{equation}
\begin{equation}
\varepsilon^{f_{xy}}_B(\bbox{p}_f)=-\chi^r{\rm sgn}~(p_x
p_y)|\Delta_b^{f_{xy}}(\bbox{p}_f)|\left(\sqrt{1-{\cal
D}\sin^2\frac{\Phi}{2}}\cos\varphi\pm\sqrt{{\cal
D}}\sin\frac{\Phi}{2}\sin\varphi\right)\enspace . \label{ecdf1q}
\end{equation}
\rrule
These branches are asymmetric with respect to the inversion
$p_y\to-p_y$.

\begin{figure}
\centerline{ \rotate[r]{
\epsfxsize=0.49\textwidth{\epsfbox{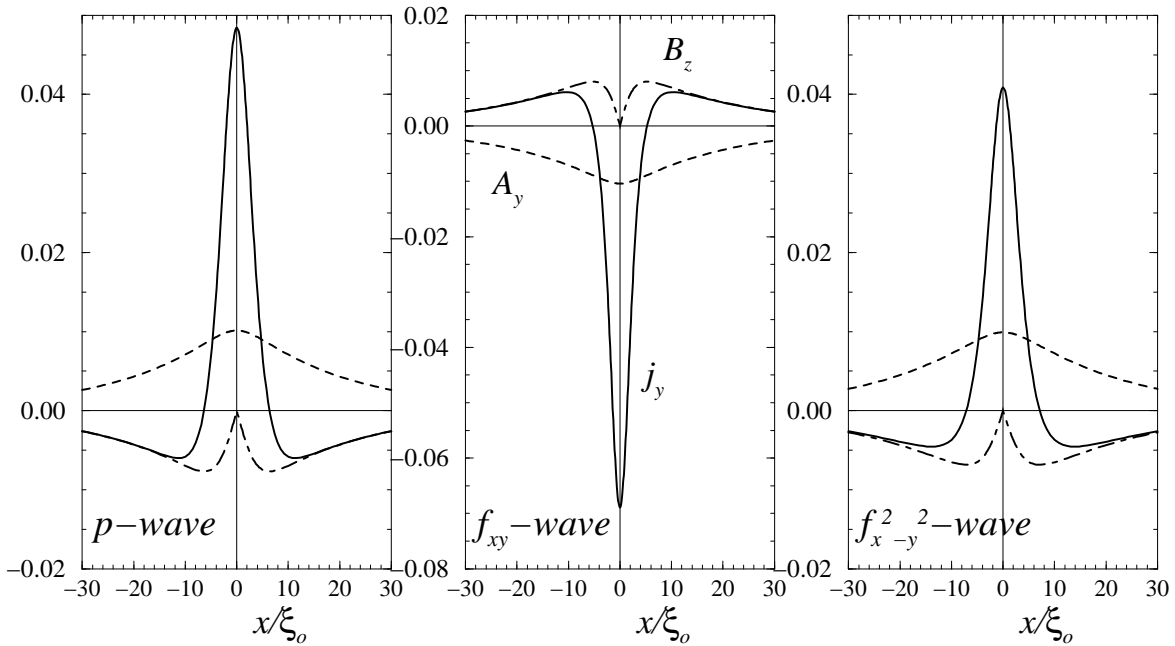}}} } \caption[]{
Spatial dependence of the surface current $j_y$, the vector
potential $A_y$ and the surface magnetic field $B_z$. The
penetration depth is set to 20 $\xi_0$ and the unit of the current
is $e v_f N_f \Delta_b$.} \label{SurfaceCurrent}
\end{figure}

The surface current, the vector potential, and the induced magnetic
field at an impenetrable barrier separating an
$\eta_1(\bbox{p}_f)+ i \eta_2(\bbox{p}_f)$ (left side) and an
$\eta_1(\bbox{p}_f)- i \eta_2(\bbox{p}_f)$ (right side)
superconductor are given in Figure \ref{SurfaceCurrent}. They
decay in the depth of the superconductor on a scale given by the
penetration depth. For the $p$-wave case there is
an agreement of our results with those in Ref.\ \onlinecite{matsum99}.
If instead the right superconductor would have
been calculated with the same parity as the left one the
currents, the vector potential, and the magnetic field would be
reversed in the right half-space.

As was shown in the preceding subsection, the order
parameters vary significantly from their bulk forms in the
vicinity of the interface. This raises a question of the
applicability of the approach used just above and based on a
simple model of spatially constant anisotropic order parameters.
We now report the results taking into account spatial
variations of the order parameters. As will be seen, they display a
good qualitative agreement with Eqs.(\ref{ecds})-(\ref{ecdf1q}).
Quantitative distinctions, however, take place\cite{n1}.

The surface pair-breaking modifies the low-energy chiral bound
states in a simple way. Spatial profiles of the order parameters
near impenetrable surfaces can be taken into account analytically
in the expressions for the energies, by replacing the order
parameters at the surface by effective ones. In general, we have
$\phi_{1,2}(x)\ne 0$ and the expressions for the effective surface
order parameters are quite cumbersome. We get a simple answer if
the surface influence retains $\phi_1(x)=\phi_2(x)=0$. The last
condition holds, in particular, for the p-wave pairing (see Fig.
\ref{OrderParameters}) and we represent here the results for this
simplest case (the other cases are considered in the Appendix).

Taking advantage of the approach of Refs.
\onlinecite{bbs97,bar00} (see also Appendix), we obtain the
bound state energies in close vicinity of the nodes of
$\Delta^{p}_2(\bbox{p}_f)$:
\begin{equation}
\varepsilon^{p}_B(\bbox{p}_f)= \pm\chi{\rm sgn}({\rm v}_x
\delta\varphi)|\Delta^{p}_{21,eff}(\bbox{p}_f,\pm0)| \enspace .
\label{spb}
\end{equation}
Here and below the upper (the lower) sign corresponds to the right
(the left) half-space; $\delta\varphi$ is the deviation of $\varphi$
from the direction of a node $\Delta^{p}_2(\bbox{p}_f)=0$ (that is
at $\varphi=0$ or $\varphi=\pi$ for the p-wave). The effective surface order
parameters are defined as
\begin{equation}
|\Delta_{ij,eff}(\bbox{p}_f,\pm0)|= \frac{\displaystyle
\int\limits_0^{\pm\infty}|\Delta_{i}(\bbox{p}_f,x)|e^{
\mp\frac{2}{|{\rm v}_x|}\int\limits^{x}_{0}
|\Delta_{j}(\bbox{p}_f,x')|dx'}dx}{\displaystyle\int\limits_0^{\pm\infty}
e^{\mp\frac{2}{|{\rm v}_x|}\int\limits^{x}_{0}
|\Delta_{j}(\bbox{p}_f,x')|dx'}dx} \label{espbr1} \enspace .
\end{equation}
Eqs.(\ref{spb}), (\ref{espbr1}) should be compared with
Eq.(\ref{ebpc}), generalizing it to spatially dependent order
parameters.
\end{multicols}
\begin{multicols}{2}

The transformation of the spectrum of surface bound states into
one of interface states due to finite transparency of the tunnel
junction,  i.e. for ${\cal D} \ll 1$, can be described analytically taking the surface
pair-breaking into account if
$\phi_i(x)=0$ (see Appendix). For instance, for the
symmetric junction ($\tilde\chi=1$) we get the following result in
vicinity of a momentum direction $\bbox{p}_{f,0}$, for which
$\varepsilon_B( \bbox{p}_{f,0})\Big\vert_{{\cal D}=0}=0$:
\lrule
\begin{equation}
\varepsilon^{p}_B(\bbox{p}_f)=\pm\left[\Bigl(
\varepsilon^{p}_B(\bbox{p}_f)\Big\vert_{{\cal D}=0}\Bigr)^2+ {\cal
D}\cos^2\left( \frac{\displaystyle \Phi}{\displaystyle2}\right)
\left(\frac{2}{|{\rm
v}_x|}\int\limits_0^{+\infty}e^{-\frac{2}{|{\rm v}_x|}
\int\limits^{x}_{0}|\Delta^{p}_{1}(\bbox{p}_f,x')|dx'}dx\right)^{-2}
\right]^{1/2} \enspace , \label{eds1}
\end{equation}
In the case of opposite chiralities ($\tilde\chi=-1$) we obtain
\begin{equation}
\varepsilon^{p}_B(\bbox{p}_f)=
\varepsilon^{p}_B(\bbox{p}_f)\Big\vert_{{\cal D}=0}\pm \sqrt{{\cal
D}}\cos\left( \frac{\displaystyle \Phi}{\displaystyle2}\right)
\left(\frac{2}{|{\rm
v}_x|}\int\limits_0^{+\infty}e^{-\frac{2}{|{\rm v}_x|}
\int\limits^{x}_{0}|\Delta^{p}_{1}(\bbox{p}_f,x')|dx'}dx\right)^{-1}
\enspace . \label{eds2}
\end{equation}

\begin{figure}
\centerline{
\rotate[r]{\epsfxsize=0.69\textwidth{\epsfbox{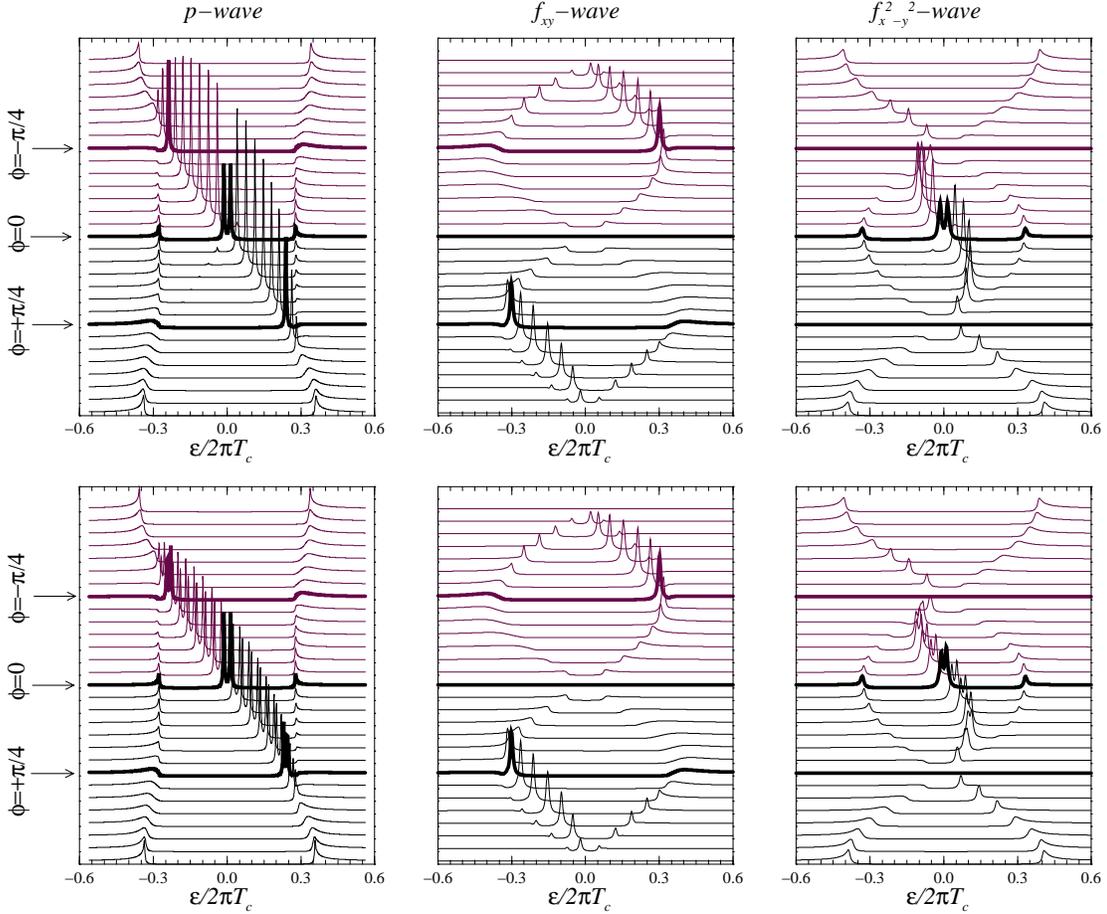}}}}
\caption[]{ The local angular resolved surface density of states
shown (left to right) for the $p$-wave, the $f_{xy}$-wave and the
$f_{x^2-y^2}$-wave junctions. The transparency ${\cal D}=0.01$ and
the phase difference is zero. The top row are all symmetric
junctions ($\tilde{\chi}=+1$) while the bottom row are junctions
with opposite chiralities ($\tilde{\chi}=-1$). The DOS is given in
the interval $\varphi=-\pi/2$ to $\varphi=\pi/2$ in equally spaced
steps. } \label{Densityofstates}
\end{figure}
\rrule

In figure \ref{Densityofstates} we show
the numerically computed local interface DOS, using the
self-consistently determined self-energies.
The angle-resolved DOS in the left superconductor is defined as
\begin{equation}
N(\varphi,\varepsilon)=-\frac{1}{2} N_f \mbox{Im}\bigg\lbrace
\mbox{Tr} \lbrack \hat \tau_3 \hat g(\varphi,0_-;\varepsilon)
\rbrack \bigg\rbrace
\end{equation}
and calculated at zero phase difference with one and the same
transparency ${\cal D}=0.01$ for all momentum directions. Here and
below we use the normalization condition $\hat{g}^2=-1$ for the
quasiclassical Green's function. The DOS in the right
superconductor are for these symmetric junctions simply related to
the DOS in the left by substituting $\varepsilon^r=
-\tilde{\chi}\cdot\varepsilon^l$. Qualitatively, peak positions
are well described by Eqs.(\ref{ebpc}) and (\ref{ebsc}), obtained
in the tunneling limit. The fine structure of the peaks are due to
a finite transmission. The density of states for junctions with
bound states dispersing in $\bbox{p}_f$ and having
$\varepsilon_B=0$ at $\varphi=0$, i.e. for the $p$-wave and for
the $f_{x^2-y^2}$-wave pairings, show a split DOS, by an amount
$\sqrt{\cal D}$ at $\varepsilon=0$ (in accordance with
Eqs.(\ref{ecds}) and (\ref{ecdsq}); see also (\ref{eds1}) and
(\ref{eds2})). At finite angles $\varphi$ the DOS show the
predicted dependence on the relative chirality of the two
superconductors. For the $\tilde{\chi}=+1$ junction, since
$\varepsilon^r_{B,{\cal D}=0}= -\varepsilon^l_{B,{\cal D}=0}$,
there is only a very small spectral weight in the DOS from the
bound state $\varepsilon^r_{B,{\cal D}}$ induced in the left
superconductor through the junction in the tunneling limit. For
the $\tilde{\chi}=-1$ junction we have $\varepsilon^r_{B,{\cal
D}=0}= \varepsilon^l_{B,{\cal D}=0}$. Hence, there is almost equal
weight in the two states split around $\varepsilon_{B,{\cal
D}=0}(\varphi)$. The split again $\propto\sqrt{{\cal D}}$ (see
Eq.(\ref{eds2})). For small angles $(\varphi\approx 0)$ the
$p$-wave and the $f_{x^2-y^2}$-wave superconductors lead to
qualitatively similar DOS. For the $f_{xy}$-wave superconductors
the chiral branches take on low energies only for momentum
directions close to the nodes of the order parameter and the main
part of the low-energy spectral weight belongs to glancing
trajectories. The $f_{xy}$-wave with $\tilde\chi=-1$ does not show
splitting if $\Phi=0$, in accordance with Eq.(\ref{ecdf1q}).

\section{Low temperature anomaly of the Josephson current}

\subsection{The Josephson current in quantum point contacts}

Consider the Josephson current across junctions with chiral
interface states first assuming spatially constant order
parameters. For a quantum point contact we can find the current in
symmetric junctions (with $\tilde\chi=1$)
as\cite{fur90,fur191,been191,been291} $j=2e\sum \limits_a
\frac{\displaystyle d\varepsilon_{B}^{a}}{\displaystyle
d\Phi}n_f(\varepsilon_{B}^{a}) = -2 e
\sum\limits_{\varepsilon_{B}^{a}>0} \frac{\displaystyle
d\varepsilon_{B}^{a}}{\displaystyle d\Phi}\tanh\frac{\displaystyle
\varepsilon_{B}^{a}}{\displaystyle 2T}$. Here the spectrum is
presumably even and the sum is taken over different channels.
Making use of the relationship $\frac{\displaystyle
d\varepsilon_B}{\displaystyle d\Phi}=\frac{\displaystyle
-|\Delta|^2{\cal D}\sin\Phi}{\displaystyle 4\varepsilon_B}$, which
easily follows from Eqs.(\ref{ecds})-(\ref{ecdf1}) for all
pairings we discuss, leads to the Josephson current
\begin{equation}
j=\frac{\displaystyle e|\Delta|^2{\cal D}\sin\Phi}{\displaystyle
2\varepsilon_B}\tanh\frac{\displaystyle\varepsilon_{B}}{\displaystyle
2 T} \label{qpc}
\end{equation}
for one channel in a quantum point contact.

Eq.(\ref{qpc}) is applicable also to contacts between isotropic
s-wave superconductors. In this case, however, the bound state
energies are mainly on the order of $\Delta$. They can take
low-energy values only in highly transparent junctions and for
phase differences $\Phi$ in a narrow vicinity of $\pi$. There are
well-known specific features of the Josephson current manifested
in conventional superconductors with
micro-constrictions\cite{kulom77}. For the phase difference
$\Phi\approx\pi$, however, $\sin\Phi$ in Eq.(\ref{qpc}) is very
small as well, precluding the low-temperature features of the
Josephson critical current we discuss below. By contrast, under
certain conditions chiral bound states can take low energy values
at any phase difference $\Phi$. For the $p$-wave and the
$f_{x^2-y^2}$-wave order parameters, as well as for the
three-dimensional $E_{1g}$-type of pairing, this is the case, in
particular, in tunnel junctions (${\cal D}\ll 1$) for
quasiparticles with momenta aligned almost (or exactly) parallel
to the interface normal $\varphi\approx 0,\pi$.

For a nonzero energy and sufficiently low temperatures, when $T
\ll\varepsilon_B $, we get from Eq.(\ref{qpc}) the
zero-temperature value of the Josephson current
$j=\frac{\displaystyle e|\Delta_0|^2{\cal D}\sin\Phi
}{\displaystyle 2\varepsilon_B}$. Here $\Delta_0$ is the
zero-temperature order parameter. Hence, the zero-temperature
critical current is much greater for channels with
$\varepsilon_B\ll|\Delta_0|$, as compared with channels, where
$\varepsilon_B\sim |\Delta_0|$, for given $\Phi$, ${\cal D}$ and
$|\Delta_0|$.

If $\varepsilon_B\ll T \ll T_c$, which can be
satisfied only by low-energy states, we find from Eq.(\ref{qpc})
that the Josephson current can vary substantially in the
low-temperature region being inversely proportional to the
temperature: $j=\frac{\displaystyle e|\Delta_0|^2{\cal
D}\sin\Phi}{\displaystyle 4T}$.

To be specific, consider the $p$-wave order parameter and
$\varepsilon_B=\Delta_b^p\sqrt{\displaystyle {\cal
R}\sin^2\varphi+{\cal D}\cos^2\frac{\displaystyle
\Phi}{\displaystyle 2}}$. Here $\varphi$ is the parameter
characterizing the quantum state in the channel. The bound state
energy is small, i.e. $\varepsilon_B\ll\Delta_b^p$,  for example
if $|\varphi|\ll 1$ and ${\cal D}\ll 1$. Then the zero-temperature
current is quite large and equal to either $\frac{\displaystyle
e\Delta_{b,0}^{p}{\cal D}\sin\Phi }{\displaystyle 2|\varphi|}$
when ${\rm max}\left(\frac{\displaystyle T}{ \displaystyle
\Delta_b^p}, \sqrt{\displaystyle{\cal
D}}\left|\cos\frac{\displaystyle \Phi}{\displaystyle
2}\right|\right)\ll |\varphi|\ll 1$ or to
$e\Delta_{b,0}^{p}\sqrt{{\cal D}}{\rm
sgn}\left(\cos\frac{\displaystyle \Phi}{\displaystyle
2}\right)\sin\frac{\displaystyle \Phi}{\displaystyle 2}$  if
$\frac{\displaystyle T}{\displaystyle \Delta_b^p}\ll |\varphi|\ll
\sqrt{\displaystyle{\cal D}}\left|\cos \frac{\displaystyle
\Phi}{\displaystyle 2}\right|\ll 1$. In the last case the
Josephson current in the tunnel junction turns out to be
proportional to $\sqrt{\cal D}$, in contrast to the linear
dependence on ${\cal D}$  of the conventional tunneling current.
As we see, the low-temperature value of the Josephson current
depends on the dimensionless parameter $\sqrt{{\cal
D}}|\Delta||\cos(\Phi/2)|/T$ containing the transparency. This
parameter appears from the expression for the energy of the
interface bound states when comparing it with the temperature. At
sufficiently low temperatures the parameter is large even if
${\cal D}\ll 1$.

It follows from Eqs.(\ref{ecds})-(\ref{ecdf1}) that the condition
$\varepsilon_B(\bbox{p}_f)\ll\Delta(\bbox{p}_f)$ is satisfied for
any phase difference for the tunnel junctions with $\varphi$ near
the nodes at $\varphi_0\approx0, \pi$ of the chiral bound states
for the $p$-wave, the $f_{x^2-y^2}$-wave and the three-dimensional
$E_{1g}$ order parameters. For the $f_{xy}$-wave order parameter
the above condition takes place only for glancing trajectories,
which contribute negligibly  to the Josephson current. For the
three-dimensional $E_{2u}$ type of pairing the nodes of the bound
states, not coinciding with the nodes of the order parameter, lie
at $\varphi^t_0=\pm\pi/4,\pm 3\pi/4$. For this reason the $E_{2u}$
pairing (\mbox{UPt$_3$}) has a contribution from the low-energy
chiral surface bound states to the Josephson current that turns
out to be especially sensitive to the momentum dependence of the
tunneling probability, ${\cal D}(\bbox{p}_{f})$. This contrasts
the $E_{1g}$ pairing as well as the $p$- and $f$-wave
superconductors. The sensitivity is associated with the fact that
the transparency of a barrier quickly diminishes with increasing a
deviation of the momentum direction from the surface normal
(unless the barrier is sufficiently thin). Thus, the transmission
can become quite small for momenta with $\varphi^t_{0}=\pm\pi/4$
or $\pm3\pi/4$. For example, a model for the tunneling barrier
with a uniform probability distribution within an acceptance cone
about the interface normal, and zero outside the cone, the effect
considered is entirely absent unless the acceptance cone contains
(or its boundary is very close to)
$\varphi^t_{0}=\pm\pi/4,\pm3\pi/4$. For a narrow
Gaussian-distribution model the tunneling at
$\varphi^t_{0}=\pm\pi/4,\pm3\pi/4$ is not strictly zero, although
exponentially small. For a sufficiently thin and high barrier,
however, the transparency is approximately proportional to
$(\bbox{n}\cdot\bbox{p}_{f})^2$ and then the contribution to the
Josephson current from the low-energy chiral surface states
becomes important even for the $E_{2u}$-triplet pairing. In the
last case the  particular $\bbox{p}_{f}$-dependence of the
tunneling probability is ${\cal D}(\bbox{p}_{f})={\cal D}_0\cos^2
\varphi$ for two-dimensional models with a cylindrical Fermi
surface and ${\cal D}(\bbox{p}_{f})={\cal
D}_0\sin^2\theta\cos^2\varphi$ for three-dimensional models with a
spherical Fermi surface.

\subsection{The critical current in classical tunnel junctions}

The low-temperature anomalous behavior of the Josephson current
described above is quite similar to what was theoretically found
for the case of $d$-wave superconductors in the presence of
zero- and low-energy surface and interface
states\cite{tk96,bbr96,rb98}. However, the similarity takes place
only in considering one channel, which is appropriate to  quantum
point contacts. In classical junctions, where quasiparticles have
various momentum directions, the parameter $\varphi$ appears in
the expression for the Josephson current as an integration
variable in averaging over the Fermi surface. If the bound states
are substantially dispersive, manifesting strong dependence of
their energy on the momentum direction, the averaging can
noticeably weaken the low-temperature deviations of the Josephson
current from its conventional behavior.

The eventual result can be easily understood qualitatively, if
one notices that the integration of the $1/|\varphi|$-term (see
the preceding subsection) over the interval $\left({\rm
max}\left(\frac{\displaystyle T}{\displaystyle \Delta_b^p},
\sqrt{\displaystyle {\cal D}}\left|\cos\frac{\displaystyle \Phi}{
\displaystyle 2}\right|\right), \varphi_c\right)$ (where
$\varphi_c\ll 1$) leads to logarithmic low-temperature dependences
on the temperature or the transparency. Below we present a more
careful analysis of the low-temperature features of the critical
current.

For a classical symmetric junction (with $\tilde\chi=1$) between
quasi-two-dimensional superconductors we get instead of Eq.(\ref{qpc}):
\begin{equation}
j_x=\int\limits_{-\frac{\pi}{2}}^{+\frac{\pi}{2}}d\varphi
\frac{\displaystyle e{\rm v}_fN_f{\cal D}(\varphi)\cos\varphi\sin{\Phi}
|\Delta(\varphi)|^2
}{\displaystyle2\varepsilon_B(\varphi)}\tanh\left(\frac{
\displaystyle\varepsilon_B(\varphi)}{\displaystyle2T}\right) .
\label{jxsc}
\end{equation}

Assuming the logarithmic functions mentioned above to be large,
one can use a logarithmic approximation, considering only a part
of the integral in Eq.(\ref{jxsc}), which contains the
zero-level crossing of the chiral branches $\bbox{p}_{f,0}$. Then
for a symmetric tunnel junction between quasi-two-dimensional
superconductors with cylindrical Fermi surfaces we get from
Eq.(\ref{jxsc})
\lrule
\begin{equation}
j_x\approx\!\!\!\sum_{\stackrel{\varphi_0}{\bbox{p}_{f,x}>0}}e{\rm
v}_fN_f{\cal D}(\varphi_0)\cos\varphi_0\sin{\Phi}
|\Delta(\varphi_0)|^2
\int\limits_{\varphi_0}^{\varphi_0+\delta\varphi_c}d\varphi\frac{\displaystyle
1}{\displaystyle\varepsilon_B(\varphi)}\tanh\left(\frac{
\displaystyle\varepsilon_B(\varphi)}{\displaystyle2T}\right) ,
\label{jxs}
\end{equation}
where $\varepsilon_B(\varphi)$ is defined by Eqs.(\ref{ecds}) and
(\ref{ecdt}), $\varepsilon_B(\varphi)\Big\vert_{{\cal D}=0}$ is to
be evaluated to linear order in $\delta\varphi$. $\delta\varphi_c$
is a cut-off parameter. The sum is taken over those $\varphi_0$,
which correspond to $\bbox{p}_{f,x}>0$. For the $p$- or the
$f_{x^2-y^2}$-wave pairing only $\varphi^{p,f}_0=0$ satisfies the
condition. In the case of three-dimensional superconductors with
the $E_{2u}$- or the $E_{1g}$-pairings, an additional integration
$\int\limits_0^{\pi/2}\sin^2\theta d\theta$ has to be carried out
in Eq.(\ref{jxsc}), where ${\cal D}$, $|\Delta|$ and
$\varepsilon_B$, depend on both $\varphi$ and $\theta$. For the
$E_{2u}$ and the $E_{1g}$ one should consider variations
$\delta\varphi$ from $\varphi^t_0=\pm\pi/4$ and $\varphi^{s}_0=0$
respectively.

The contribution to the Josephson current from low-energy bound
states in the tunnel junctions can be also calculated analytically
taking into account the surface pair-breaking. Indeed, from
Eqs.(\ref{eds1}), (\ref{ed1}) we easily get $\frac{\displaystyle
d\varepsilon_B}{\displaystyle d\Phi}=\frac{\displaystyle
-|\tilde{\Delta}|^2{\cal D}\sin\Phi}{\displaystyle
4\varepsilon_B}$, where the effective order parameter
$|\tilde{\Delta}(\bbox{p}_{f},0)|$ near the nodes of the bound
states is defined in the right half space
\begin{equation}
\frac{\displaystyle 1}{\displaystyle
|\tilde{\Delta}(\bbox{p}_{f},0)|}= \frac{\displaystyle
2}{\displaystyle |{\rm v}_{f,x}(\bbox{p}_{f})|}
\int\limits_0^{\infty} e^{-\frac{ 2}{ {\rm v}_{f,x}
(\bbox{p}_{f})}\int\limits_0^{x_1} |\Delta(\bbox{p}_{f},
x_2)|\sin\left(\eta(\bbox{p}_{f},x_2)
-\phi(\bbox{p}_{f},x_2)\right)dx_2 }dx_1 \enspace \label{|effdel|}
\enspace ,
\end{equation}
\rrule
and analogously in the left half space.
Hence, spatial dependence of the order parameters introduces in this case
the only modification in Eq.(\ref{jxs}): the order parameter
$|\Delta(\varphi)|$ should be replaced there by the effective
surface order parameter $|\tilde{\Delta}(\varphi)|$.

For temperatures $\sqrt{{\cal D}(\bbox{p}_{f,0})}|\tilde{\Delta}(
\bbox{p}_{f,0})| |\cos(\Phi/2)|\ll T\ll \Delta_b\delta\varphi_c<
T_c$ one can put $\tanh\left(\varepsilon_B(\varphi)/2T\right)\approx 1$
near the upper limit of the integral in Eq.(\ref{jxs}) and use
Eqs.(\ref{ecds}) and (\ref{ecdt}) for
$\varepsilon^{p,f_{x^2-y^2}}_B(\varphi),\varepsilon^{s,t}_B(\theta,\varphi)
\propto\delta\varphi$. Then we find a logarithmic temperature
dependence of the integral near the lower limit $\varphi_{0}$ in
Eq.(\ref{jxs}). Qualitatively, the temperature is a lower cut of
the integral, but for fixing numerical factors we checked its
asymptotic behavior numerically. Thus, at sufficiently low
temperatures, when contributions to the current from
quasiparticles with energies the order of $|\Delta(\bbox{p}_{f})|$
already take their zero-temperature value, the Josephson current
associated with the low-energy part of the chiral bound states and
described by Eq.(\ref{jxs}) can still grow logarithmically with
the decreasing temperature. As a result, the total Josephson
current can be represented for the low-temperature interval
$\sqrt{{\cal D}(\bbox{p}_{f,0})}|\tilde{\Delta}(\bbox{p}_{f,0})|
|\cos(\Phi/2)|\ll T\ll T_c$ as
\begin{equation}
j_x(T)=aj_{c,0}\ln\left(\frac{\displaystyle b\Delta_b}{\displaystyle
T}\right )
\sin\Phi
\label{jost}
\enspace ,
\end{equation}
where $j_{c,0}=LeN_f{\rm v}_f{\cal D}_0\Delta_b$ ($L^p=2/3$,
 $L^{f_{x^2-y^2}}\approx0.45$, $L^{f_{xy}}=2/5$, $L^s=1/9$, $L^t=2/21$),
$\Delta_b=\Delta^{p,f_{x^2-y^2},f_{xy},s,t}_b$, $a$ and $b$ are
constants the order of unity. In our analytical results here and
below the transparency is taken for high and thin potential
barriers as it was represented in the end of the preceding
subsection. Further, disregarding surface pair-breaking, we
calculate analytically a total Josephson current and find at low
temperatures after the comparison with Eq.(\ref{jost}) $a^p=3/2$,
$b^p\approx0.69$, $a^{f_{x^2-y^2}}\approx2.23$,
$b^{f_{x^2-y^2}}\approx0.44$, $a^s=3/2$, $b^s\approx0.23$,
$a^t=3/4\sqrt{2}$, $b^t\approx 0.95$. In neglecting not only a
surface pair-breaking (that is a spatial dependence of the order
parameter), but the presence of any surface bound states at all,
one can use spatially constant (bulk) values of Green's functions
for the calculation of the Josephson current. This oversimplified
approach leads for the symmetric tunnel junction to the Josephson
critical current $j_{c,0}$. As the argument of the logarithmic
function in Eq.(\ref{jost}) is supposed to be quite large (for the
logarithmic approximation to be valid), the Josephson current
$j_x(T)$ can noticeably exceed $j_{c,0}$.

For temperatures $T\ll \sqrt{{\cal D}_0}\Delta_b|\cos(\Phi/2)|$
the lower cut of the integral is associated with the transparency
of the barrier and we find from Eq.(\ref{jxs})
\begin{equation}
j_x(0)=aj_{c,0}\ln\left(\frac{\displaystyle d}{\displaystyle
\sqrt{{\cal D}_0}|\cos(\Phi/2)|}\right) \sin\Phi \label{josd}
\enspace.
\end{equation}
The logarithmic function in Eq.(\ref{josd}) is supposed to take
large values as well. Disregarding surface pair breaking, we
find $d^p\approx1.21$, $d^{f_{x^2-y^2}}\approx0.78$,
$d^s\approx1.21$, $d^t\approx1.88$.

Based on Eqs.(\ref{ecdsq})-(\ref{ecdf1q}), one can easily get the
Josephson current in junctions with opposite chiralities. We do
not present here the corresponding analytical results in detail,
since there is no unconventional low-temperature increase in the
critical current in the case $\tilde\chi=-1$. In disregarding
surface pair breaking the Josephson current in tunnel junction at
$T\ll T_c$ in the case $\tilde\chi=-1$ is written as
\begin{equation}
j_x=uj_{c,0}\sin\Phi\enspace, \label{och}
\end{equation}
where $u^p=9/8$,\, $u^{f_{xy}}=5/3$,\,
$u^{f_{x^2-y^2}}\approx1.95$,\, $u^s=9/8$,\, $u^t\approx0.142$.
The ``reference'' current $j_{c,0}$ is taken here the same as in
Eqs.(\ref{jost}) and (\ref{josd}) in order to make clear that the
Josephson current in the case of opposite chiralities is of the
conventional order of magnitude without any logarithmic
enhancement. The smaller numerical factor $u^t$ is associated with
the lower transparency for $\varphi^t_0=\pm\pi/4$ as compared with
$\varphi^{p,f_{x^2-y^2},s}_0=0$. Surface chiral bound states in
the whole range of subgap energies form the current in the
particular case, even at low temperatures. One can show that the
low-energy chiral quasiparticle states lead to a current which is
reduced compared to Eq.(\ref{och}) by the  small cut-off parameter
$\delta\varphi_c$ and it does not matter whether the surface pair
breaking is taken into account or not.

Another reason for choosing the ``reference'' current $j_{c,0}$ in
the case $\tilde\chi=-1$ same as for $\tilde\chi=1$, is that in
neglecting both the surface pair breaking and the presence of
Andreev surface bound states, and simply using the bulk
expressions for the Green's functions, the tunnel Josephson
critical current across the domain wall ($\tilde\chi=-1$)
vanishes. This in sharp contrast with Eq.(\ref{och}). For a
vanishing Josephson current one needs, for example, different
projections of the orbital angular momentum, i.e. different values
of $l_z$, of the Cooper pairs in the left and the right
superconductors together with a conservation of $l_z$ in the
tunneling process. The current vanishes, for instance, if $l_z$ is
along the interface normal and the interface is symmetric with
respect to the rotations around the normal\cite{amb74,mrs88}. As
we demonstrate in Eq.(\ref{och}), if $l_z$ is along an axis
parallel to the interface, the interface influence results in
interface bound states and the possibility of a nonzero tunneling
of Cooper pairs. Eq.(\ref{och}) applies to tunnel junctions.
Effects of high transparency can also lead to finite critical
current\cite{yip93}.

On account of the bound states, the Green's function depends on the distance
from the surface even when disregarding a spatial dependence of the order
parameter. There is an important qualitative difference between the surface and
the bulk values of the propagators as a function of momentum direction. In
particular, the angle $\varphi$ is an important part of the phase not only for
the chiral order parameter, but for the bulk value of the off-diagonal components
of the Green's function as well. Since we consider the $z$-axis to be parallel
to the surface, the surface breaks the conservation of the projection of the
angular momentum. For this reason the angle $\varphi$ drops out from the phase
in the surface value of the off-diagonal components. This can be obtained on
account of the continuity of the quasiclassical Green's function, taken for
incoming and outgoing trajectories (with the angles $\varphi$ and $\pi-\varphi$
respectively) at the impenetrable surface.  Thus, on account of the surface
effects, the Josephson tunnel current across the domain wall is of the
conventional value (see Eq.(\ref{och})).

\begin{figure}
\centerline{
\rotate[r]{\epsfxsize=0.59\textwidth{\epsfbox{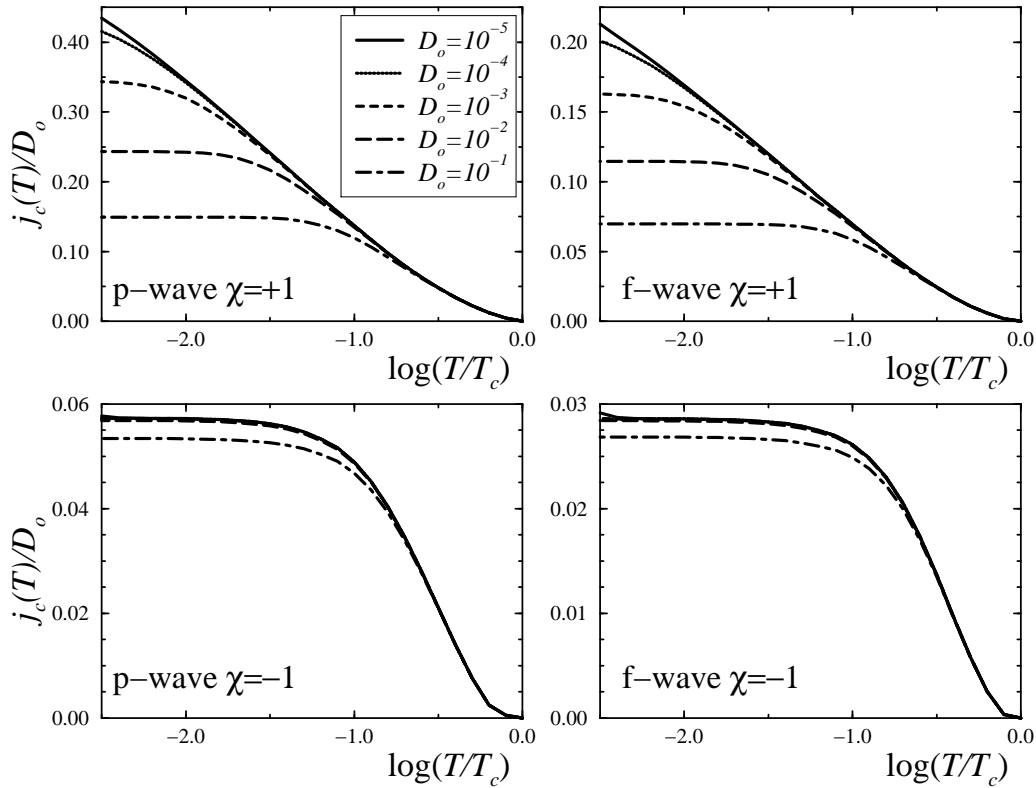}}}}
\caption[]{ The temperature dependence of the critical current
divided by the maximal transparency ${\cal D}_0$ plotted for
different values of ${\cal D}_0$. The junctions are modelled with
the angle dependent transparency ${\cal D}(\varphi)= {\cal D}_0
\exp (- 16 \sin^2 \varphi)$. To the left we show two coupled
p-wave superconductors and to the right two $f_{x^2-y^2}$-wave
ones. } \label{CriticalCurrents}
\end{figure}

The logarithmic behavior of the critical current found above for
classical symmetric junctions with $\tilde\chi=1$, is observable
only at sufficiently low temperatures and at low junction
transparencies when the logarithmic approximation can be
justified. This is demonstrated in Figure \ref{CriticalCurrents},
where we show the result of numerical calculations for the
temperature dependence of the critical current for different
values of barrier transparency.
\end{multicols}
\begin{multicols}{2}

As seen for the ${\tilde\chi}=+1$ junctions, the term ${\cal
D}_0\ln {\cal D}_0$ is present in the tunneling limit in the
dependence of the zero-temperature critical current on the
transparency (see Eq.(\ref{josd})). Further, the low-temperature
range, where the critical current linearly depends on $\ln T$ (see
Eq.(\ref{jost})), becomes well pronounced only for transparencies
less than $10^{-2}$. The finite transparency efficiently cuts off
the logarithmic temperature dependence of the critical current.
The lower the transparency, the larger the low-temperature range.
The characteristic temperature, below which the transparency
influence becomes noticeable, is proportional to $\sqrt {{\cal
D}}$, again in agreement with our analytical results.

\begin{figure}
\centerline{
\rotate[r]{\epsfxsize=0.59\textwidth{\epsfbox{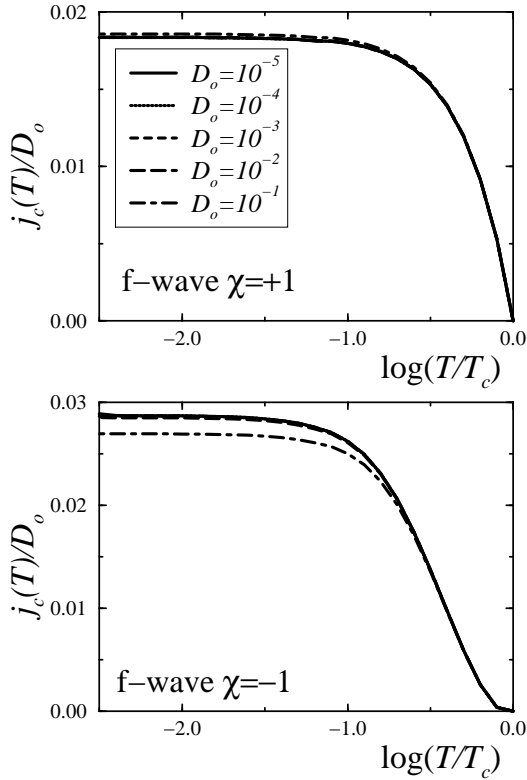}}}}
\caption[]{ The temperature dependence of the critical current
calculated as in Fig.\ \ref{CriticalCurrents} but for the
$f_{xy}$-wave superconductor.} \label{CriticalCurrents_f0}
\end{figure}

In Fig.\ \ref{CriticalCurrents_f0} we show the analogous result
for the $f_{xy}$-wave superconductor. Since the low-energy
contribution from the bound states is associated in this case only
with glancing trajectories, there are no anomalies in the critical
current.

\subsection{Effects of broadening of the bound states}

Broadening of the chiral low-energy bound states can substantially
modify the low-temperature behavior of the critical current found
above. Even a small broadening $\gamma\ll \Delta_b$  have a
profound influence on the low-energy parts of chiral branches. We
take a small broadening into account in the pole-like term of the
retarded quasiclassical Green's function, where we simply replace
the factor $1/[\varepsilon-\varepsilon_B(\bbox{p}_f)]$ with
$1/[\varepsilon-\varepsilon_B(\bbox{p}_f)+i\gamma]$, analogously
to what is done in\cite{bkk00}. This slightly modifies the
analysis of Ref.\ \onlinecite{bbr96} for the contribution from the
low-energy bound states to the Josephson current in tunnel
junctions. We evaluate the current in the tunneling limit assuming
the momentum independent broadening substantially greater than the
low-energy bound state and the temperature:\ $T,\sqrt{{\cal
D}_0}\Delta_b|\cos( \Phi/2)|\ll \gamma\ll\Delta_b$. The lower cut
of the integral in Eq.(\ref{jxs}) is associated in this case with
$\gamma$ rather than the temperature or the bound state energy.
Then we find for the Josephson current
\begin{equation}
j_x(0)=aj_{c,0}\ln\left(\frac{\displaystyle
w\Delta_b}{\displaystyle \gamma}\right ) \sin\Phi \label{josgamma}
\enspace .
\end{equation}
For spatially constant order parameters we get $w^p\approx0.61$,
$w^{f_{x^2-y^2}}\approx0.39$, $w^s\approx0.21$, $w^t\approx0.84$.

There are various contributions to the broadening of the bound
states. These are in particular associated with surface roughness,
non-magnetic and magnetic impurities, and with inelastic
scattering. We assume here that non-magnetic impurities dominate
the scattering and thus the broadening. Then we calculate the
effect of small scattering rates on the behavior of the Josephson
current at low temperatures using the usual t-matrix approximation
and assuming an isotropic impurity potential, $u$. The impurity
self energy is in this case
\begin{equation}
\hat \Sigma(\varepsilon)=\Gamma {\langle \hat g(\varepsilon) \rangle \over
{1-\sigma(1+\langle \hat g(\varepsilon)\rangle^2})}
\label{self_impurity}
\enspace.
\end{equation}
where the scattering strength $\sigma=\sin^2\delta_s$ ($\delta_s$
is the s-wave scattering phase-shift, $0\le
\delta_s\le\frac{\pi}{2}$) and the scattering rate, $\Gamma$,
parameterize $u$ and the impurity density $n_{imp}$
as
\begin{equation}
\sigma=\frac{(\pi N_f u)^2}{1+(\pi N_f u)^2} \qquad
{\Gamma}=\frac{\pi N_f u^2}{1+(\pi N_f u)^2}\,n_{imp} \label{tau}
\enspace .
\end{equation}

Due to the low-energy bound states, the Matsubara Green's function
can take quite large values in the low-temperature region, if the
broadening and the transparency are sufficiently small. However,
the low-energy states form only a small part of the chiral
branches, i.e. for most of the quasiparticle trajectories the
energy of the chiral states is the order of
$\Delta_b(\bbox{p}_f)$. For this reason the quasiclassical
propagator for the chiral superconductor, averaged over the Fermi
surface, $\langle g(\varepsilon)\rangle$, does not take
large values but is of the order of unity or less. This differs
greatly from the case of a 45$^\circ$-oriented d-wave
superconductor for which each trajectory has an Andreev bound state at
zero energy. There $\langle g(\varepsilon) \rangle$ can take large
values along with the pole-like term in the quasiclassical Green's
function\cite{pbbi99}.

\end{multicols}
\begin{figure}
\centerline{
\rotate[r]{\epsfxsize=0.59\textwidth{\epsfbox{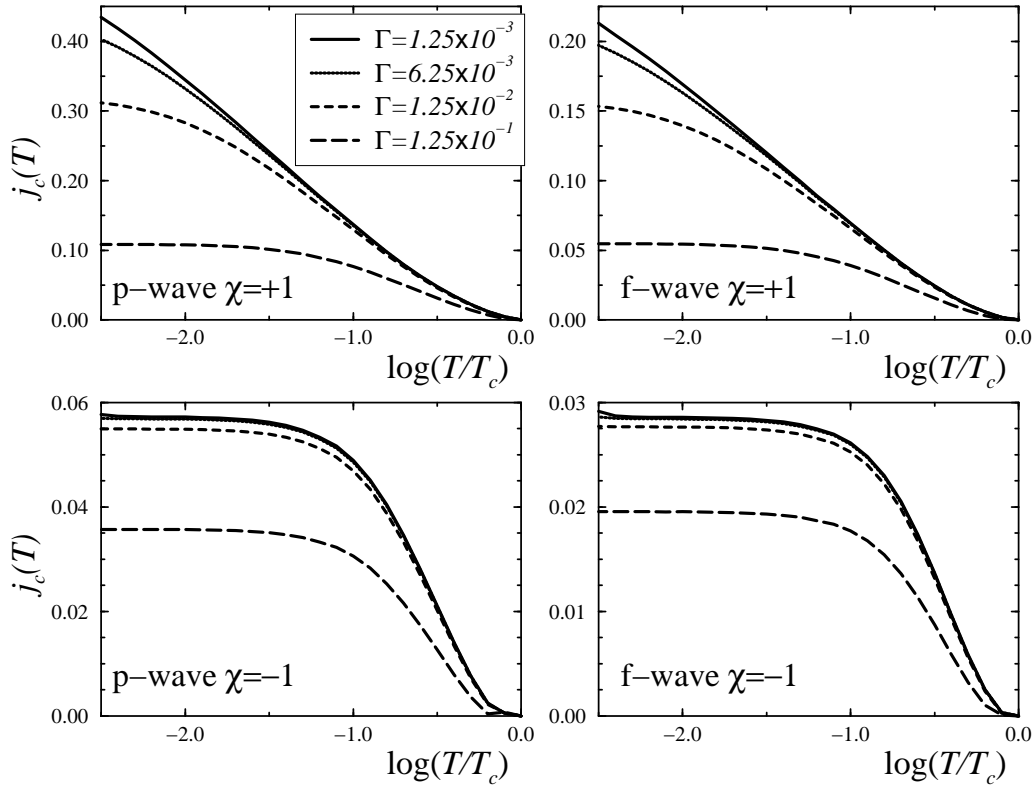}}}}
\caption[]{ The temperature dependence of the critical current for
a junction modelled with the transparency ${\cal D}(\varphi)=
{\cal D}_0 \exp (- 16 \sin^2 \varphi)$ and ${\cal D}_0=10^{-5}$.
$j_c(T)$ is calculated at different scattering rates $\Gamma
/T_c$. The scattering strength is weak and put to $\sigma=0.1$. }
\label{JC_vs_Imps}
\end{figure}
\begin{multicols}{2}

In Fig.\ \ref{JC_vs_Imps} we show the influence of impurity
effects on the temperature dependence of critical current. In
order to get a wide range of well pronounced dependence of the
critical current on the small broadening, we take the extreme tunnel
limit and put ${\cal D}_0=10^{-5}$. One can see from the Fig.\
\ref{JC_vs_Imps}, that for observing the logarithmic
low-temperature enhancement of $j_c$, one needs
superconductors of high purity.

\section{Summary}

Chiral interface Andreev bound states have been obtained and
studied above both analytically and numerically. We showed, that
the low-energy chiral states results in the low-temperature
enhancement of the Josephson current between clean chiral
superconductors in symmetric tunnel junctions. The enhancement is
more pronounced in quantum point contacts. In classical junctions
the zero-temperature current acquires an additional logarithmic
dependence on low transparency or on the broadening of the
bound states. Under the conditions considered, the Josephson
current through the domain wall does not vanish due to the bound
state contribution.

\section{Acknowledgments}
Yu.B. would like to thank T.~Kopp and J.~Mannhart for kind
hospitality during his stay in University of Augsburg, where a
part of this work was carried out. This work was supported in part
by BMBF 13N6918/1 (Yu.B.), by the Russian Foundation for Basic
Research under Grant No. 99-02-17906 (Yu.B. and A.B.) and by the
Swedish Natural Science Research Council (M.F.).

\end{multicols}
\appendix
\section*{Energies of the interface bound states}

In the presence of a quasiparticle bound state
$\varepsilon_B(\bbox{p}_f)$ the quasiclassical retarded propagator $\hat g$
has a pole at $\varepsilon=\varepsilon_B(\bbox{p}_f)$. One can
introduce the residue of the propagator $\hat g$  as
\begin{equation}
\hat{\tilde g}(\bbox{p}_f,\bbox{r};\varepsilon_B(\bbox{p}_f))=
\lim_{\varepsilon\to\varepsilon_B(\bbox{p}_f)}
\left[(\varepsilon-\varepsilon_B(\bbox{p}_f)) \hat
g(\bbox{p}_f,\bbox{r};\varepsilon)\right] \enspace ,
\label{gtilde}
\end{equation}
which is finite and satisfies the same transport equation as $\hat
g$ but completed with the relation
\begin{equation}
\big[ \hat{\tilde g}(\bbox{p}_f,\bbox{r};
\varepsilon_B(\bbox{p}_f)) \big]^2=0 \enspace , \label{mbc}
\end{equation}
rather than the normalization condition.

For calculating the bound state energies, the Eilenberger
equation for $\hat{\tilde g}$ can be solved in terms of the
following ansatz\cite{bbs97}:
\begin{eqnarray}
 \tilde{f}^{+}(\bbox{ p}_{f},x;\varepsilon_B(\bbox{p}_f))=
\tilde g  (\bbox{ p}_{f},x;\varepsilon_B(\bbox{p}_f))
exp(-i\eta(\bbox{p}_f, x))
\enspace , \nonumber \\
\tilde{f}(\bbox{ p}_{f},x;\varepsilon_B(\bbox{p}_f))= -\tilde g
(\bbox{ p}_{f},x;\varepsilon_B(\bbox{p}_f)) exp(i\eta(\bbox{p}_f,
x)) \enspace . \label{symmetry}
\end{eqnarray}

The whole number of quasiclassical equations can then be reduced
to the one scalar equation
\begin{equation}
-\frac{{\rm v}_{f,x}}{2}\partial_x\eta(\bbox{p}_f,x)+
\varepsilon_B(\bbox{p}_f)-
|\Delta(\bbox{p}_f,x)|\cos\left(\eta(\bbox{p}_f,x)-
\phi(\bbox{p}_f,x)\right)=0 \enspace , \label{pe}
\end{equation}
completed with the condition at the interface\cite{bar00}
$$
D\sin\left(\frac{\displaystyle \eta_{l,0}(\bbox{p}_{f,l})-
\eta_{r,0}(\bbox{p}_{f,r})}{\displaystyle 2}\right)
\sin\left(\frac{\displaystyle
\eta_{l,0}(\underline{\bbox{p}}_{f,l})-
\eta_{r,0}(\underline{\bbox{p}}_{f,r})}{\displaystyle 2}\right)=
\qquad\qquad\qquad\qquad\qquad\qquad
$$
\begin{equation}
\qquad\qquad\qquad\qquad\qquad\qquad\qquad
\sin\left(\frac{\displaystyle \eta_{l,0}(\bbox{p}_{f,l})-
\eta_{l,0}(\underline{\bbox{p}}_{f,l})}{\displaystyle 2}\right)
\sin\left(\frac{\displaystyle \eta_{r,0}(\bbox{p}_{f,r})-
\eta_{r,0}(\underline{\bbox{p}}_{f,r})}{\displaystyle 2}\right)
\label{t}
\end{equation}
and the asymptotic conditions in the right and the left
superconductors
\begin{equation}
{\rm v}_{f,x}(\bbox{p}_{f})
\sin\left(\eta_{\infty}(\bbox{p}_{f})-\phi_{\infty}(\bbox{p}_{f})
\right) {\rm sgn}x>0 \enspace , \qquad x\rightarrow\pm\infty
\enspace . \label{asr}
\end{equation}

Eqs.(\ref{pe})-(\ref{asr}) are valid both for singlet
superconductors and for triplet ones with a $(0,0,d_z)$
order parameter. Eq.(\ref{t}) connects
solutions of Eq.(\ref{pe}) with momenta $\bbox{p}_{f,l}$,
$\bbox{p}_{f,r}$ of incoming quasiparticles from the left and
the right sides of the interface with the momenta
$\underline{\bbox{p}}_{f,l}$, $\underline{\bbox{p}}_{f,r}$ of
reflected ones. For specular reflection, the momentum
parallel to the interface is conserved, i.e.,
$\bbox{p}^\parallel_{f,l}=\underline{\bbox{p}}_{f,l}^\parallel
=\bbox{p}^{\parallel}_{r,f}=\underline{\bbox{p}}^{\parallel}_{f,r}$.
In the limit of impenetrable wall Eq.(\ref{t}) reduces to a
continuity condition for $\eta_{0}(\bbox{p}_{f})$ taken for
incoming and outgoing momenta along a quasiparticle trajectory.

Eq.(\ref{pe}) can be easily solved for spatially independent order
parameters. Thus, for a symmetric junction between the $p$-wave
superconductors (with any chiralities) one gets
\begin{equation}
\left\{\begin{array}{ll}
\eta_l(\bbox{p}_{f,l})\!=\!\Phi_l+\!\chi_l\varphi\!-\arccos\!\left(\frac{\displaystyle
\varepsilon_B(\bbox{p}_{f,l})}{\Delta_b^p}\right),&
\eta_l(\underline{\bbox{p}}_{f,l})\!=\!\Phi_l+\chi_l(\pi-\varphi)+\arccos\!\left(
\frac{\displaystyle\varepsilon_B(\bbox{p}_{f,l})}{\Delta_b^p}\right)\\
\eta_r(\underline{\bbox{p}}_{f,r})\!=\!\Phi_r\!+\!\chi_r\varphi\!+\!\arccos\!\left(\frac{
\displaystyle\varepsilon_B(\bbox{p}_{f,l})}{\Delta_b^p}\right),&
\eta_r(\bbox{p}_{f,r})\!=\!\Phi_r+\chi_r(\pi-\varphi)-\arccos\!\left(
\frac{\displaystyle\varepsilon_B(\bbox{p}_{f,l})}{\Delta_b^p}\right)
\end{array}
\right. \label{sol}
\end{equation}

Substitution of Eq.(\ref{sol}) into Eq.(\ref{t}) results in
Eq.(\ref{ecds}) for the spectrum of the bound states in the case
of identical chiralities and in Eq.(\ref{ecdsq}) for opposite
chiralities. Similar derivations with the other types of pairing
lead to Eqs.(\ref{ecds})-(\ref{ecdf1}) and
Eqs.(\ref{ecdsq})-(\ref{ecdf1q}).

Calculations of surface and interface chiral bound state energies,
taking into account the spatial profile of the order parameter,
are carried out for the low-energy states with momenta close to
nodes (or to low-energy minima) $\bbox{p}_{f,0}$ of the chiral
branches $\varepsilon_B(\bbox{p}_{f})$. Consider, for
example, the $p$-wave order parameter. We note, that
$\Delta^p_{2}(\bbox{p}_f,x)=0$ for two particular momentum
directions: $\varphi^p_0=0$ and $\pi$. On the other hand,
$\varphi^p_0=0$ and $\pi$ correspond to the incoming and the outgoing
momenta in a reflection event where
$\Delta^p_{1}(\bbox{p}_f,x)$ changes its sign. Hence, the
zero-energy surface bound states take place near an impenetrable
surface for these particular quasiparticle trajectories.
For the $p$-wave order parameter the solution of Eq.(\ref{pe})
for the zero-energy states is
$\eta_p^{(0)}=\Phi+\frac{\displaystyle\pi}{\displaystyle2}{\rm
sgn}( x)$. At an impenetrable surface a chiral branch crosses
the zero energy with a slope,
which can be found for a spatially dependent order parameter.
Indeed, considering a small deviation
$\delta\varphi=\varphi-\varphi^p_0$ from any of the two trajectories,
we linearize Eq.(\ref{pe}) and get the following first-order
corrections to $\eta_p^{(0)}$:
\begin{equation}
\eta_p^{(1)}(x,\bbox{p}_f)=-\frac{\displaystyle2}{\displaystyle
{\rm v}_x}
\int\limits_x^{\pm\infty}\left(\varepsilon^p_B(\bbox{p}_f)
-\chi^{r(l)}{\rm sgn}(p_yx')|\Delta^p_{2}(\bbox{p}_f,x')|\right)e^{
\mp\frac{2}{|{\rm v}_x|}\int\limits_{x}^{x'}
|\Delta^p_{1}(\bbox{p}_f,x'')|dx''}dx' \enspace . \label{esar}
\end{equation}
Here and below the upper and the lower signs correspond to half-spaces $x>0$
and $x<0$ respectively.

At the impenetrable wall $\eta(\bbox{p}_f,x=0)=\eta_p^{(0)}+
\eta_p^{(1)}(x=0,\bbox{p}_f)$ is one and the same for
incoming and outgoing momentum directions. Taking this into
account we find Eq.(\ref{spb}) for the bound state energies
close to $\varphi^p_0$ for p-wave pairing.

Calculations for the $E_{1g}$-case are very close to the $p$-wave
pairing since the orbital parts of the order parameters contain
identical dependence on $p_x$ and $p_y$. Thus, we get in this case
$\varepsilon^{s}_B(\bbox{p}_f)= \pm\chi{\rm sgn}({\rm v}_x
\delta\varphi)|\Delta^{s}_{21,eff}(\bbox{p}_f,\pm0)|$ with the same
notations as in Eq.(\ref{spb}).

Analogous consideration can be carried out for the $E_{2u}$
triplet pairing. Then zero energy bound states occur at
four particular momentum directions with
$\varphi^t_0=\pm\pi/4,\pm3\pi/4$ and this at any fixed value of $\theta$.
The solution of Eq.(\ref{pe}) for \,$\varphi=\varphi^t_0$\, is\,
$\eta_t^{(0)}=\Phi+\frac{\displaystyle\pi}{\displaystyle2}
\left(1+ {\rm sgn}( x\chi p_yp_z)\right)$. Corrections to
$\eta_t^{(0)}$ linear in small deviations $\delta\varphi$ from
$\varphi^t_0$, are described by the equation
\begin{equation}
\eta_t^{(1)}(x,\bbox{p}_f)=-\frac{\displaystyle2}{\displaystyle|{\rm
v}_x|}
\int\limits_x^{\pm\infty}\left(\varepsilon^t_B(\bbox{p}_f){\rm
sgn}({\rm v}_x) -\chi^{r(l)}{\rm
sgn}(x'\delta\varphi)|\Delta^t_{1}(\bbox{p}_f,x')|\right)e^{
\mp\frac{2}{|{\rm v}_x|}\int\limits_{x}^{x'}
|\Delta^t_{2}(\bbox{p}_f,x'')|dx''}dx' \label{etar} \enspace .
\end{equation}

Then in a narrow region of the nodes of $\Delta^t_1(\bbox{p}_f)$ we
obtain for an impenetrable wall $\varepsilon^t_B(\bbox{p}_f)=\pm
\chi{\rm sgn}({\rm v}_x
\delta\varphi)|\Delta^t_{12,eff}(\bbox{p}_f,\pm0)| \label{deff}$
where $\delta\varphi$ is the deviation of $\varphi$ from
the direction of a node: $\Delta^{t}_1(\bbox{p}_f)=0$.

If the equality $\phi_i(x)=0$, assumed above, is not satisfied,
the result for the effective surface order parameter becomes
more cumbersome. First, one should find the function
$\eta^{(0)}(x)$ satisfying the equation
\begin{equation}
\frac{{\rm v}_{f}}{2}\partial_x\eta^{(0)}(x) +
|\Delta_1(\bbox{p}_{f,0},x)|\cos\left(\eta^{(0)}(x)-
\phi_1(x)-\Phi\right)=0 \enspace
\end{equation}
and the asymptotic conditions
$\eta^{(0)}(\pm\infty)=\Phi+(\pi/2){\rm sgn~}x$.

Then, introducing the notation $\zeta_i(x)=\eta^{(0)}(x)-\phi_i(x)-\Phi$,
we obtain, for example, for the $f_{x^2-y^2}$-wave pairing
the following result:
\begin{equation}
\varepsilon^{f_{x^2-y^2}}_B(\bbox{p}_f)= \chi^{r(l)}{\rm
sgn~v}_y\frac{\displaystyle
\int\limits_0^{\pm\infty}|\Delta_{2}(\bbox{p}_f,x)|\sin
\zeta_2(x)e^{-\frac{2}{|{\rm v}_x|}\int\limits^{x}_{0}
|\Delta_{1}(\bbox{p}_f,x')|\sin
\zeta_1(x')dx'}dx}{\displaystyle\int\limits_0^{\pm\infty}
e^{-\frac{2}{|{\rm v}_x|}\int\limits^{x}_{0}
|\Delta_{1}(\bbox{p}_f,x')|\sin \zeta_1(x')dx'}dx} \label{ef45r}
\enspace .
\end{equation}

Let now the transmission be
finite, but sufficiently small. Assume $\phi_i(x)=0$, which takes
place, for instance, for the $p$-wave pairing.
In vicinities of
the momentum directions $\bbox{p}^p_{f,0}$, where
$\Delta^p_{2,b}( \bbox{p}^p_{f,0})=0$ ( that is
$\varphi=\varphi^p_0$ ) we get
$\eta_p^{(0)}=\Phi+\frac{\displaystyle\pi}{\displaystyle2}{\rm
sgn}( x)-{\rm sgn}(x{\rm v}_x)\varepsilon^{p}_B(\bbox{p}^p_{f,0})/
|\Delta^p_{1,b}(\bbox{p}^p_{f,0})|$.
Eq.(\ref{esar}) can be then generalized to the presence of nonzero
transparency:
$$
\eta_p^{(1)}(x,\bbox{p}_f)=-\frac{\displaystyle2}{\displaystyle{\rm
v}_x}
\int\limits_x^{\pm\infty}\Biggl(\varepsilon^p_B(\bbox{p}_f)-\frac{\displaystyle
|\Delta^p_{1}(\bbox{p}_f,x')|}{\displaystyle|\Delta^p_{b}(\bbox{p}_f)|}
\varepsilon^{p}_B(\bbox{p}^p_{f,0})- \qquad\qquad\qquad
\qquad\qquad\qquad
$$
\begin{equation}
\qquad\qquad\qquad \qquad\qquad\qquad -\chi^{r(l)}{\rm sgn}({\rm
v}_xx'\delta\varphi)|\Delta^p_{2}(\bbox{p}_f,x')|\Biggr)e^{
\mp\frac{2}{|{\rm v}_x|}\int\limits_{x}^{x'}
|\Delta^p_{1}(\bbox{p}_f,x'')|dx''}dx' \label{esard}
\enspace.
\end{equation}

Analogously, near momentum directions $\bbox{p}^t_{f,0}$, for
which $\Delta^t_{1,b}(\bbox{p}^t_{f,0})=0$ (that is
$\varphi_0=\pm\frac{\pi}{4},\pm\frac{3\pi}{4}$) we get
$\eta_t^{(0)}=\Phi+\frac{\displaystyle\pi}{\displaystyle2}
\left(1+ {\rm sgn}( x\chi p_yp_z)\right)-{\rm sgn}(x{\rm
v}_x)(\varepsilon^{t}_B(\bbox{p}^t_{f,0})/
|\Delta^t_{2,b}(\bbox{p}^t_{f,0})|$. Then in the presence of a nonzero
transparency we get
$$
\eta_t^{(1)}(x,\bbox{p}_f)=-\frac{\displaystyle2}{\displaystyle{\rm
v}_x}
\int\limits_x^{\pm\infty}\Biggl(\varepsilon^t_B(\bbox{p}_f)-\frac{\displaystyle
|\Delta^t_{2}(\bbox{p}_f,x')|}{\displaystyle|\Delta^t_{b}(\bbox{p}_f)|}
\varepsilon^{t}_B(\bbox{p}^t_{f,0})- \qquad\qquad\qquad
\qquad\qquad\qquad
$$
\begin{equation}
\qquad\qquad\qquad \qquad\qquad\qquad -\chi^{r(l)}{\rm sgn}({\rm
v}_xx'\delta\varphi)|\Delta^t_{1}(\bbox{p}_f,x')|\Biggr)e^{
\mp\frac{2}{|{\rm v}_x|}\int\limits_{x}^{x'}
|\Delta^t_{2}(\bbox{p}_f,x'')|dx''}dx' \label{etard}
\enspace .
\end{equation}

For finding the bound state energies one should insert solutions
from Eq.(\ref{esard})   (Eq. (\ref{etard})) into the condition at
the interface Eq.(\ref{t}), taking account of the first
non-vanishing corrections in transparency or in $\delta\varphi$.
For a symmetric tunnel junction, when the order parameters on both
sides coincide for every momentum direction and, in particular, have
identical chiralities, this expansion leads to Eqs.(\ref{eds1}) for the
$p$-wave pairing as well as for the $E_{1g}$ pairing. For the
$E_{2u}$ type of pairing we get in this case
\begin{equation}
\varepsilon^t_B(\bbox{p}_f)=\pm\left[\Bigl(
\varepsilon^t_B(\bbox{p}_f)\Big\vert_{{\cal D}=0}\Bigr)^2+ {\cal
D}\cos^2\left( \frac{\displaystyle \Phi}{\displaystyle2}\right)
\left(\frac{2}{|{\rm
v}_x|}\int\limits_0^{+\infty}e^{-\frac{2}{|{\rm v}_x|}
\int\limits^{x}_{0}|\Delta^t_{2}(\bbox{p}_f,x')|dx'}dx\right)^{-2}\right]^{1/2}
\enspace . \label{ed1}
\end{equation}
For the domain wall ($\tilde\chi=-1$) the spectrum for the
$p$-wave pairing is described by Eq.(\ref{eds2}). The same
expression is valid as well for the $E_{1g}$ pairing. For the
$E_{2u}$ type of pairing we get in the case of opposite
chiralities
\begin{equation}
\varepsilon^t_B(\bbox{p}_f)=
\varepsilon^t_B(\bbox{p}_f)\Big\vert_{{\cal D}=0}\pm \sqrt{{\cal
D}}\sin\left( \frac{\displaystyle \Phi}{\displaystyle2}\right)
\left(\frac{2}{|{\rm
v}_x|}\int\limits_0^{+\infty}e^{-\frac{2}{|{\rm v}_x|}
\int\limits^{x}_{0}|\Delta^t_{2}(\bbox{p}_f,x')|dx'}dx\right)^{-1}
\enspace . \label{ed2}
\end{equation}

\begin{multicols}{2}



\end{multicols}

\begin{references}
\bibitem{n1}
If the surface pair-breaking depends noticeably on
crystal-to-interface orientations, it would induce qualitative
distortions of the orientational dependencies of spectra of
Andreev interface states. This is definitely the case for $d$-wave
superconductors\cite{bs00}, but, in particular, not present for
the $p$-wave pairing, since there is no surface-to-crystal
anisotropy in the surface influence on the $p$-wave order
parameter.
\end{references}
\end{document}